\documentclass[12pt,preprint]{aastex}

\newcommand{\lsim}{\raise0.3ex\hbox{$<$}\kern-0.75em{\lower0.65ex\hbox{$\sim$}}}
\newcommand{\gsim}{\raise0.3ex\hbox{$>$}\kern-0.75em{\lower0.65ex\hbox{$\sim$}}}
\newcommand{\propsim}{\raise0.3ex\hbox{$\propto$}\kern-0.75em{\lower0.65ex\hbox{$\sim$}}}

\begin{document}
    
\title{Joining radio with X-rays: A revised model for SN 1993J}

\author{C.-I. Bj\"ornsson\altaffilmark{1}}
\altaffiltext{1}{Department of Astronomy, AlbaNova University Center, Stockholm University, SE--106~91 Stockholm, Sweden.}
\email{bjornsson@astro.su.se}

\begin{abstract}
A joint analysis is done of the radio and X-ray observations of SN 1993J. It is argued that neither synchrotron cooling behind the forward shock nor thermal cooling behind the reverse shock is supported by observations. In order for adiabatic models to be consistent, a reinterpretation of the radius of the spatially resolved VLBI-source is needed during the first few hundred days. Instead of reflecting the position of the forward shock, it is then associated with the expansion of the Rayleigh-Taylor unstable region emanating from the contact discontinuity. Although observations imply a constant ratio between the energy densities in magnetic fields and relativistic electrons, they do not appear to scale individually with the thermal energy density behind the forward shock; rather, in adiabatic models, the evolution of the magnetic field strength is best understood as scaling inversely with the supernova radius.
\end{abstract}

\keywords{radiation mechanisms: non-thermal --- stars: mass-loss --- supernovae: general --- supernovae: individual (SN 1993J)}

\section{Introduction}
SN 1993J is one of the most well-observed supernovae, detected in the radio and all the way to soft $\gamma$-rays. In addition, VLBI-observations spatially resolved the radio source early on. The interaction of the supernova ejecta and the circumstellar medium, due to the wind from progenitor star, creates shock waves, which are usually taken to be described by the self-similar solutions derived by \cite{che82a}. The standard model ascribes both the radio and X-ray emission to come from the region in between the forward and reverse shocks \citep{che82b}. It is therefore expected that an analysis of the radio and X-ray observations should give complementary information regarding the physical conditions in the shocked region. Although there have been a number of papers addressing the physical implications of various aspects of the radio or X-ray observations from SN 1993J \citep[e.g.,][]{suz93, s/n95, f/b98, per01, mio01}, only a few have attempted the synergy that a joint analysis could give \citep[e.g.,][]{fra96}.

Many of the observed properties of SN 1993J stand out in comparison with other supernovae. Although the observations strongly indicate that the initial X-ray emission was dominated by free-free emission \citep{zim94, tan94, lei94}, it proved hard to model the observations with a single component evolving with time; for example, the soft X-ray declined quite slowly while the harder X-ray showed a more standard behavior \citep{cha09}. Intriguing as this may be, the most distinct features show up in the radio regime. 

The VLBI-observations revealed a conspicuous dynamical change at around $300$\, days \citep[e.g.,][]{bar02, bie11}, when the expansion velocity changed from being nearly constant to a rapidly decreasing phase. The spatially resolved observations also showed that the brightness temperature was substantially below that expected for a homogeneous synchrotron source in which magnetic fields and relativistic electrons were in equipartition. Associated with the changing dynamical behavior was a distinct change in the variation of the synchrotron self-absorption frequency. During the initial, almost constant velocity phase, the value of the synchrotron self-absorption frequency decreased slowly before transiting to a more rapid decline. This later phase is in line with expectations from the standard model and similar to that in many other radio supernovae. If the evolution of the self-absorption frequency is related to the dynamics, spatially resolved observations should not be needed in order to detect such a behavior. Other well-observed, but spatially unresolved, radio supernovae have not given any clear indications of a similar behavior, which makes SN 1993J quite unique in this respect. 

There has been a substantial amount of observations published after most of the models for SN 1993J were proposed; in particular, in the radio by \cite{wei07} and a compilation of X-ray data by \cite{cha09}. The first aim of the present paper is to critically discuss some aspects of these early models to see how they fare in a comparison with the later observations. Although the starting point is the radio properties of SN 1993J, it is shown in Section\,\ref{sect2} that a joint discussion of the radio and X-ray observations provides important insights to a physical understanding of the shocked region. It is concluded that the earlier models need modifications and/or additions in order to account simultaneously for both the radio and X-ray emission. Section\,\ref{sect3} presents a toy model, which remedies some of these shortcomings. It is argued in Section\,\ref{sect4} that the Rayleigh-Taylor instability at the contact discontinuity can give a physical content to the toy model and, hence, make it a viable model for the radio as well as the X-ray emission. It is also shown how such a model can provide a framework in which several of the observed properties of SN 1993J, not discussed in detail in this paper, can be understood. A summary of the main conclusions follows in Section\,\ref{sect5}.

\section{Modeling the radio observation of SN 1993J} \label{sect2}

A homogeneous, spherically symmetric, expanding synchrotron source is often sufficient to adequately model the radio observations of supernovae. The detailed observations of SN 1993J, including angularly resolved VLBI measurements \citep[e.g.,][]{mar97, bar02, bie03}, made it possible to test several of the simplifying assumptions included in the standard model. The radio emission is thought to originate from behind the forward shock. For most radio supernovae only two independent observables can be obtained (in addition to the spectral index) from observations, namely, the synchrotron self-absorption frequency ($\nu_{\rm abs}$) and the corresponding spectral flux ($F(\nu_{\rm abs})$). In order to deduce the magnetic field strength and source radius some assumption needs to be made regarding the energy density of the relativistic electrons with respect to that in the magnetic fields.

The angularly resolved observations provided a third observable for SN 1993J. Although the assumed sphericity was confirmed by observations \citep{bie03}, the derived brightness temperature was lower than expected for equipartition between magnetic fields and relativistic electrons. In a homogeneous source, this necessitates the energy density of the magnetic fields to be much larger than that in relativistic electrons. As shown by \cite{f/b98} and \cite{per01} the magnetic field implied is so strong that synchrotron cooling affects the value and temporal evolution of $\nu_{\rm abs}$. This then also accounts for the initially unusually slow temporal decrease of $\nu_{\rm abs}$. The reason is that the importance of cooling decreases with time and, hence, the column density of relativistic electrons decreases less rapidly than expected in the standard model.

Although the standard model with a large deviation from equipartition between magnetic fields and relativistic electrons gives a good fit to the observations, there are a few implications which may suggest the need to revise some of its basic tenets. Since these were not elaborated on in either \cite{f/b98} or \cite{per01}, they are discussed in turn below. When a quantitative comparison is needed, the results in \cite{f/b98} will be used.

For a steady wind from the progenitor star, the thermal energy density behind the forward shock varies with time ($t$) as $t^{-2}$. It is often assumed that the energy density in the magnetic fields scales with the thermal energy density so that the magnetic field strength $B \propto t^{-1}$. In SN 1993J, \cite{f/b98} deduced that the magnetic energy density normalized to the thermal energy density is $\epsilon_{\rm B} \approx 0.14$, while the corresponding energy density of relativistic electrons ($\epsilon_{\rm e}$) was several hundred times smaller. An alternative scaling of the magnetic field corresponds to the situation when its energy density varies inversely with the radiating surface, i.e., for a spherically symmetric source $B\propto R^{-1}$, where $R$ is its radius. Such a scaling implies a non-constant value for $\epsilon_{\rm B}$.

The mechanism responsible for the amplification of the magnetic field behind the shock is not well understood. A correct description of its scaling relation based on observations is likely to constrain possible scenarios; for example, in order to distinguish between the $t^{-1}$ and $R^{-1}$ scaling relations, a rapidly decreasing velocity of the forward shock is helpful. In the self-similar solutions of \cite{che82a}, this corresponds to a low value of $n$ ($R\propto t^{(n-3)/(n-2)}$), where $n$ is the power law index of the density structure of the supernova ejecta. The angularly resolved observations of SN 1993J shows that after an initial phase of almost constant expansion velocity, the outer radius of the radio source makes a distinct transition to a strongly decelerating phase ($n\approx 6-7$) after a few hundred days. Hence, the late time evolution of the radio emission of SN 1993J offers a possibility to throw light on the scaling of the magnetic field strength. Unfortunately, in the modeling by \cite{f/b98} the error bars for the $B$-values deduced from a $\chi^2$-fitting procedure increase substantially in the decelerating late phase as compared to the early, nearly constant velocity phase. Although, formally, a $R^{-1}$ scaling is preferred over a $t^{-1}$ scaling, the error bars are large enough that the latter cannot be excluded. The reason for this increase of the error bars can be understood as follows.

When synchrotron cooling is important, the optically thin spectral flux can be written
\begin{equation}
	\nu F(\nu) \propto R^2 v_{\rm sh} U_{\rm e}(\gamma),
	\label{eq1}
\end{equation}
Where $v_{\rm sh}$ is the velocity of the forward shock in the standard model and $U_{\rm e}(\gamma)$ is the energy density of relativistic electrons with Lorentz factor $\gamma$. The value of $\gamma$ to be used in equation (\ref{eq1}) is that corresponding to emission at $\nu$ (i.e., $\gamma \propto (\nu/B)^{1/2}$). For a distribution of electron energies given by $N(\gamma) \propto \gamma^{-p}$ for $\gamma > \gamma_{\rm min}$ and $p>2$, $U_{\rm e}(\gamma) \approx U_{\rm e} (\gamma/\gamma_{\rm min})^{2-p}$, where $U_{\rm e}$ is the total energy density in relativistic electrons. \cite{f/b98} used $p=2.1$ so that the variation of $U_{\rm e}(\gamma)$ with $\gamma$ is small; in order to simplify the notation, $U_{\rm e}(\gamma) \approx U_{\rm e}$ will be used below. Since $R \approx v_{\rm sh}t$,
\begin{equation}
	\nu F(\nu) \propto v_{\rm sh}^3 t^2 U_{\rm e}.
	\label{eq2}
\end{equation}
This shows that in the early, nearly constant velocity phase, the observed flat light curves imply $U_{\rm e} \propto t^{-2}$. In the decelerating phase, \cite{wei07} find that all the well observed optically thin light curves have $F(\nu) \propto t^{-0.7}$ up to approximately $3100$\,days (when a rapid achromatic decline of all the light curves sets in). For $n\approx 6-7$, this again implies  $U_{\rm e} \propto t^{-2}$. Although it is not possible to distinguish between a $R^{-1}$ and $t^{-1}$ scaling relation during the nearly constant velocity phase, the decelerating phase clearly shows that in a cooling scenario the energy density of relativistic electrons scales with the thermal energy behind the shock. Furthermore, since the variation of the energy density of relativistic electrons is a direct reflection of the observed light curves, the small error bars for the former deduced in \cite{f/b98}  is due to the small scatter in the light curves.

The light curves are independent of the magnetic field strength as long as  synchrotron cooling is important. Instead, the variation of $B$ can be deduced from the changing values of $\nu_{\rm abs}$
\begin{equation}
	\nu_{\rm abs}^3 \propto U_{\rm e} U_{\rm B} \Delta r_{\rm cool},
	\label{eq3}
\end{equation}
where $p=2$ has been used. Here, $U_{\rm B}=B^2/8\pi$ is the energy density in the magnetic field and $\Delta r_{\rm cool}$ is the distance electrons move behind the shock before they lose most of their energy. The cooling time is $\propto B^{-2} \gamma^{-1}$, which leads to
\begin{equation}
	\nu_{\rm abs}^3 \propto \frac{U_{\rm e} v_{\rm sh}}{\gamma_{\rm abs}}.
	\label{eq4}
\end{equation}
Since $\gamma_{\rm abs}$ is the Lorentz factor corresponding to $\nu_{\rm abs}$,
\begin{equation}
	\nu_{\rm abs} \propto (U_{\rm e} v_{\rm sh})^{2/7} B^{1/7}.
	\label{eq5}
\end{equation}
During the nearly constant velocity phase, \citet{f/b98}  find that $\nu_{\rm abs} \propto t^{-0.68}$ gives a good representation of the time dependence of the self-absorption frequency. With $U_{\rm e}\propto t^{-2}$, as obtained independently from the light curves, $B \propto t^{-1}$ is implied. Again, the small error bars for the values of $B$ reflect the small scatter in the deduced values of $\nu_{\rm abs}$.

For the deceleration phase, \cite{f/b98}  find $\nu_{\rm abs} \propto t^{-0.81}$. However, in contrast to the earlier phase, the error bars on the deduced $B$-values are substantially larger. This is not due to an increased observed scatter in the values of $\nu_{\rm abs}$, since it is similar in the two phases (the dispersion in the exponents is $\pm 0.03$). Since synchrotron cooling affects the value of $\nu_{\rm abs}$ also in the beginning of the deceleration phase, the change in its time dependence is due mainly to the changing velocity. It is seen from equation (\ref{eq5}) that $\nu_{\rm abs} \propto v_{\rm sh}^{2/7}$ for $B\propto t^{-1}$ and $\nu_{\rm abs} \propto v_{\rm sh}^{1/7}$ for $B\propto R^{-1}$. The changing behavior of $\nu_{\rm abs}$ is consistet with $B\propto t^{-1}$ (e.g., $v_{\rm sh} \propto t^{-0.25}$ for $n=6$), while $B\propto R^{-1}$ gives a too small of a change. This need for a rapidly decreasing $B$-value is in tension with the optically thin spectra and light curves, which indicate that cooling is more important than a $B \propto t^{-1}$ scaling would suggest; hence, the increased error bars for $B$.

The need for a scaling relation of $B$ slower than $t^{-1}$ is reinforced by the observations of \cite{wei07}, which show a steady power law decline of all the optically thin light curves from $\gsim 100$\,days until $\approx 3100$\,days. In the model of \citet{f/b98}, this means that if cooling was important at the beginning of the decelerating phase, it must be so also at $3100$\,days. The longest wavelength with a well observed light curve is $\lambda = 20$\,cm. This, then, gives a lower limit to the magnetic field strength, which corresponds, roughly, to a $R^{-1}$ scaling from the nearly constant velocity phase and, hence, $\epsilon_{\rm B} \,\gsim \,0.78$. In fact, a value larger than this lower limit is suggested by the variation of $F(\nu_{\rm abs})$. Together with $U_{\rm e} \propto t^{-2}$, equation (\ref{eq2}) yields
\begin{equation}
	F(\nu_{\rm abs}) \propto \frac{v_{\rm sh}^{3}}{\nu_{\rm abs}},
	\label{eq6}
\end{equation}
which shows that in the model of \cite{f/b98}, $F(\nu_{\rm abs})$ is expected to stay roughly constant (i.e., independent of $\nu_{\rm abs}$). This is indeed the case for the light curves observed by \citet{wei07} during the deceleration phase. It should be noticed that the roughly constant value of $F(\nu_{\rm abs})$ is model dependent, since it results from the combination of a specific velocity law and the assumption of cooling. Although the light curve for $\lambda = 90$\,cm is not of the same quality as those at shorter wavelengths, it is consistent with a constant value of $F(\nu_{\rm abs})$. The maximum occurs at $t \approx 2000$\,days, which results in an unrealistically large value of $B$ (i.e., $\epsilon_{\rm B} > 1$). Such a large value of $\epsilon_{\rm B}$ could be avoided by invoking a larger mass-loss rate from the progenitor star. This would increase the mass and energy requirements for the ejecta (see Section\,\ref{sect3}). Furthermore, in order to accommodate the free-free absorption, a heating mechanism of the circumstellar medium has to be found that is more efficient than that discussed in \cite{fra96}.

\subsection{Models with negligible synchrotron cooling} \label{sect2a}

A revised model for SN 1993J will be discussed in Section\,\ref{sect3}, for which synchrotron cooling is not important. The light curves in the optically thin phase are then given by $F(\nu) \approx F(\nu_{\rm abs}) (\nu_{\rm abs}/\nu)^{(p-1)/2}$. Assuming no cooling, \cite{f/b98} deduced $p=2.7$. Together with $F(\nu_{\rm abs})\,\approx$\,constant and $\nu_{\rm abs}\propto t^{-0.81}$, this shows that adiabatic models can account for the optically thin light curves ($F(\nu) \,\propsim \,t^{-0.7}$) found by \citet{wei07} in the deceleration phase. Furthermore, since \citet{bie11} found $R \,\propsim \,t^{0.80}$ for the outer radius of the radio source (i.e., $n \approx 7$), $\nu_{\rm abs}\,\propsim \,R^{-1}$, which is consistent with $B \propto R^{-1}$ but not $B \propto t^{-1}$. 

For a standard synchrotron model, $F(\nu_{\rm abs})\propto (\epsilon_{\rm e}/\epsilon_{\rm B})^{5/(p+4)}$. The rapid decline of the velocity in SN 1993J is particularly useful to estimate how much the scaling of $U_{\rm e}$ can deviate from that of $U_{\rm B} \propto R^{-2}$; for example,  $U_{\rm e}\propto t^{-2}$ implies $F(\nu_{\rm abs})\propto v_{\rm sh}^{10/(p+4)} \propto \nu_{\rm abs}^{10/(p+4)(n-3)}$, where $\nu_{\rm abs} \propto R^{-1}$ has been used. With $n\approx 7$ and $p = 2.7$, $F(\nu_{\rm abs}) \,\propsim \,\nu_{\rm abs}^{0.37}$. This implies that between $\lambda = 3.6$\,cm and $\lambda = 20$\,cm, $F(\nu_{\rm abs})$ should decline by a factor $1.9$, which is too fast to be consistent with observation \citep{wei07}. Hence, this suggests that also for $U_{\rm e}$ the scaling is closer to $R^{-2}$ than $t^{-2}$. A more detailed discussion of the scaling relations for $U_{\rm e}$ and $U_{\rm B}$ in radio supernovae will be given in a forthcoming paper.

\subsection{Constraints on the properties of the supernova ejecta and the wind from the progenitor star}\label{sect2c}
 
With a constant mass-loss rate ($\dot M$) from the progenitor star, the supernova shock has swept up a circumstellar mass ($M_{\rm su} $) at time $t$ corresponding to
\begin{equation}
	M_{\rm su} = \frac{n-2}{n-3} \dot M t \frac{v_{\rm sh}}{v_{\rm w}},
	\label{eq9}
\end{equation}
where $v_{\rm sh}/v_{\rm w}$ is the ratio of the velocities of the forward shock and the wind of the progenitor star, respectively. In the thin shell approximation, the ejecta mass behind the reverse shock is $M_{\rm ej} = ((n-4)/2)M_{\rm su}$, which can be written
\begin{equation}
	M_{\rm ej, \odot} = 10^{-2} \frac{(n-4)(n-2)}{2(n-3)}\dot M_{\rm -5} t_{\rm yr} \frac{v_{\rm sh,4}}{v_{\rm w,1}}, 
	\label{eq10}
\end{equation}
where $M_{\rm ej, \odot}$ is the ejecta mass behind the reverse shock in solar mass units, $\dot M_{\rm -5}$ is the mass-loss rate of the progenitor star in units of $10^{-5}$ solar masses per year, $t_{\rm yr}$ the time since the supernova explosion measured in years, $v_{\rm sh,4} \equiv v_{\rm sh}/10^4$\,km\,s$^{-1}$, and $v_{\rm w,1} \equiv v_{\rm w}/10$\,km\,s$^{-1}$.

The best-fit parameters obtained in \cite{f/b98}  are $\dot M_{\rm -5} = 5$, $t = 100$\,days, and $v_{\rm sh,4} = 2.2$. From equation (\ref{eq10}), this yields $M_{\rm ej, \odot} = 1.5 \times 10^{-2} (n-4)(n-2)/(n-3)$. As deduced from the spatially resolved VLBI-observations, the velocity of the outer extent of the radio emission moves with roughly constant velocity, which implies a large value for $n$. \cite{fra96} argued that the X-ray observations from SN 1993J are best accounted for by $n \approx 25$, which then corresponds to $M_{\rm ej, \odot} = 0.33$. For $n \approx 20-30$, $v_{\rm sh} = 2.2\times 10^4$\,km\,s$^{-1}$ corresponds to an ejecta velocity $v_{\rm ej} \approx 2.0 \times 10^4$\,km\,s$^{-1}$. The value of $M_{\rm ej, \odot}$ deduced from equation (\ref{eq10}) is then more than an order of magnitude larger than that obtained by \cite{woo94} for their best fit model (13B; $ 0.016 M_{\rm \odot}$ moving faster than $ 2.0 \times 10^4$\,km\,s$^{-1}$) to the optical observations of SN 1993J. This discrepancy between the deduced values of the ejecta mass is likely to be even larger, since the roughly constant velocity phase appears to last a factor of a few longer than assumed in \cite{f/b98} \citep{bar02}. Other models of the supernova explosion do not seem to be able to bridge this gap in ejecta masses; on the contrary, models discussed in \cite{s/n95} have ejecta masses at high velocities substantially smaller than those obtained by \cite{woo94}.

The implications of large ejecta masses at high velocities can also be illustrated by considering the corresponding energy. The energy contained in the ejecta at velocities larger than $v_{\rm ej}$ is
\begin{equation}
	 E_{\rm ej} (v_{\rm ej})= \frac{n-3}{2(n-5)} M_{\rm ej} v_{\rm ej}^2. 
	\label{eq11}
\end{equation}
With $n = 25$ and $t = 300$\,days \citep{bar02}, equation (\ref{eq11}) gives $ E_{\rm ej} (v_{\rm ej}= 2.0 \times 10^4 {\rm km/s}) \approx 4.4 \times 10^{51}$\,erg/s. Such large values of the energy at high velocities lie outside the range thought possible in standard core collapse supernovae and, qualitatively, are more similar to engine driven supernovae \citep{sod10}. In order to obtain physically more realistic values for $M_{\rm ej}$ and $E_{\rm ej}$, one needs to lower the values of $\dot M$ and/or $n$.

\cite{fra96} considered two qualitatively different models for the X-ray emission from SN 1993J. One model assumed cooling to be important behind the reverse shock. This results in absorption of the X-ray emission by the cold gas and, hence, initially no X-ray emission is expected to be observed from the reverse shock. As emphasized in \cite{fra96}, the most characteristic aspect of this radiative scenario is the evolution of the X-ray flux. After about a hundred days, the cold gas becomes transparent and the reverse shock starts to contribute to the observed X-ray emission. The declining X-ray light curves should then reverse and start to rise (ROSAT-band) or flatten out (ASCA-band). In the other model, cooling is assumed to be unimportant behind the reverse shock. Therefore, in this model, the observed flux should come from the forward as well as the reverse shock also during the initial phase of shock expansion. The two models imply quite different values for $\dot M_{\rm -5}$ and $n$. In the radiative case $\dot M_{\rm -5}\approx 3-5$ and $n\approx 25-30$, while in the adiabatic case $\dot M_{\rm -5}\approx 1-3$ and $n\approx 6-8$. \cite{fra96} argued that observations favored the radiative model. However, as has been discussed by \cite{cha09}, there are no indications from late time observations of a break in the X-ray light curves neither in the ROSAT-band nor the ASCA-band at around a few hundred days. Hence, this supports an adiabatic model, which has been advocated by \cite{suz93}, and is in line with the conclusion drawn above.

The density of the circumstellar medium is normally taken to be a power-law, $\rho_{\rm w} \propto r^{-s}$; for a steady wind, $s=2$. The complexities of both the radio and X-ray emission from SN 1993J have sometimes been attributed to a non-steady wind, i.e., $s \neq 2$. \cite{s/n95} modeled  the X-rays with $s < 2$ during an initial phase, which transits at around day $40$ to a phase with $s > 2$. This causes the velocity of the forward shock to decrease faster initially, while at later times slower, than for $s = 2$. Qualitatively, this expected flattening of the variation of the shock velocity with time is opposite the behavior observed by the spatially resolved VLBI-measurements of \cite{bar02}.

It is generally agreed that synchrotron self-absorption as well as free-free absorption are needed to account for the spectra and light curves of the radio emission from SN 1993J \citep{f/b98, che98, per01, mio01}. Even so, this does not give a good fit for $s=2$, since, for example, the resulting light curves are too narrow. As discussed in Section\,\ref{sect2}, this can, at least in the initial phase, be remedied by assuming a magnetic field strong enough to cause synchrotron cooling. Another possibility suggested by \cite{mio01} is that $s < 2$. Since the time variation of the free-free absorption ($\tau_{\rm ff}$) is deduced from observations, this has implications for the temperature structure of the circumstellar medium ($T_{\rm CM}$). \cite{f/b98} showed that $s = 2$ together with a declining value of $T_{\rm CM}$ ahead of the forward shock resulted in free-free absorption consistent with observations. With  $\tau_{\rm ff} \propto n_{\rm e}^2/T_{\rm CM}^{3/2}$, where $n_{\rm e} \propto \rho_{\rm w}$ is the density of thermal electrons, the value of $s$ deduced by \cite{mio01} instead implies $T_{\rm CM} \approx$ constant.

In \cite{f/b98}, the variation of $T_{\rm CM}$ was taken from self-consistent calculations of the Compton heating by the X-ray emission from the shock, while in \cite{mio01} it was treated as a free parameter. The much lower densities in the latter model implies negligible Compton heating. However, as discussed in \cite{b/l14}, the free-free absorption itself causes heating of the circumstellar medium. This then sets a lower value for  $T_{\rm CM}$. During the initial phase of radio emission from SN 1993J, when free-free absorption is most prominent, this lower limit is a factor of a few larger than the value  $T_{\rm CM} = 7 \times 10^4$\,K deduced by \cite{mio01}. 

As will be discussed further in Section\,\ref{sect4}, a distinguishing feature of the radio emission from SN 1993J is the low brightness temperature. This is another observational constraint that is not directly encompassed by models with  $s < 2$,  since the value of $s$ should not affect the brightness temperature. Together, this shows that there is no compelling evidence for $s \neq 2$ neither from X-ray nor radio observations.

It was argued in \cite{b/l14} that a self-consistent calculation of the temperature of the circumstellar medium, assuming heating only by the observed free-free absorption, gives a fit at least as good as the one obtained in \cite{f/b98}. The main free parameter is the density of the circumstellar medium, i.e., the mass-loss rate of the progenitor star. The deduced value lies in the range $\dot M_{\rm -5} = 0.8 - 1.0$ for $v_{\rm w,1} = 1$. The low density of the circumstellar medium implied by such a mass-loss rate makes additional heating mechanisms unlikely. The VLBI-observations show the velocity, interpreted as that of the forward shock, after about a year, to vary as $n \approx 7$. An important assumption underlying the claim of a good fit in \cite{b/l14} was that this velocity-law also applies to the initial phase, i.e., during the first few hundred days of shock expansion. If this were the case, the observed VLBI-velocity during this phase cannot correspond to that of the forward shock. The implications of such a model are discussed in the next section. It may be noticed that $\dot M_{\rm -5} = 0.8 - 1.0$ and $n \approx 7$ give values of the ejecta mass at high velocities consistent with that of model 13B in \cite{woo94}. Incidentally, these values are also close to those deduced by \cite{suz93} from an analysis of the first few weeks of X-ray observations.

\section{A revised model for SN 1993J}\label{sect3}

\begin{figure}
\epsscale{0.75}
\plotone{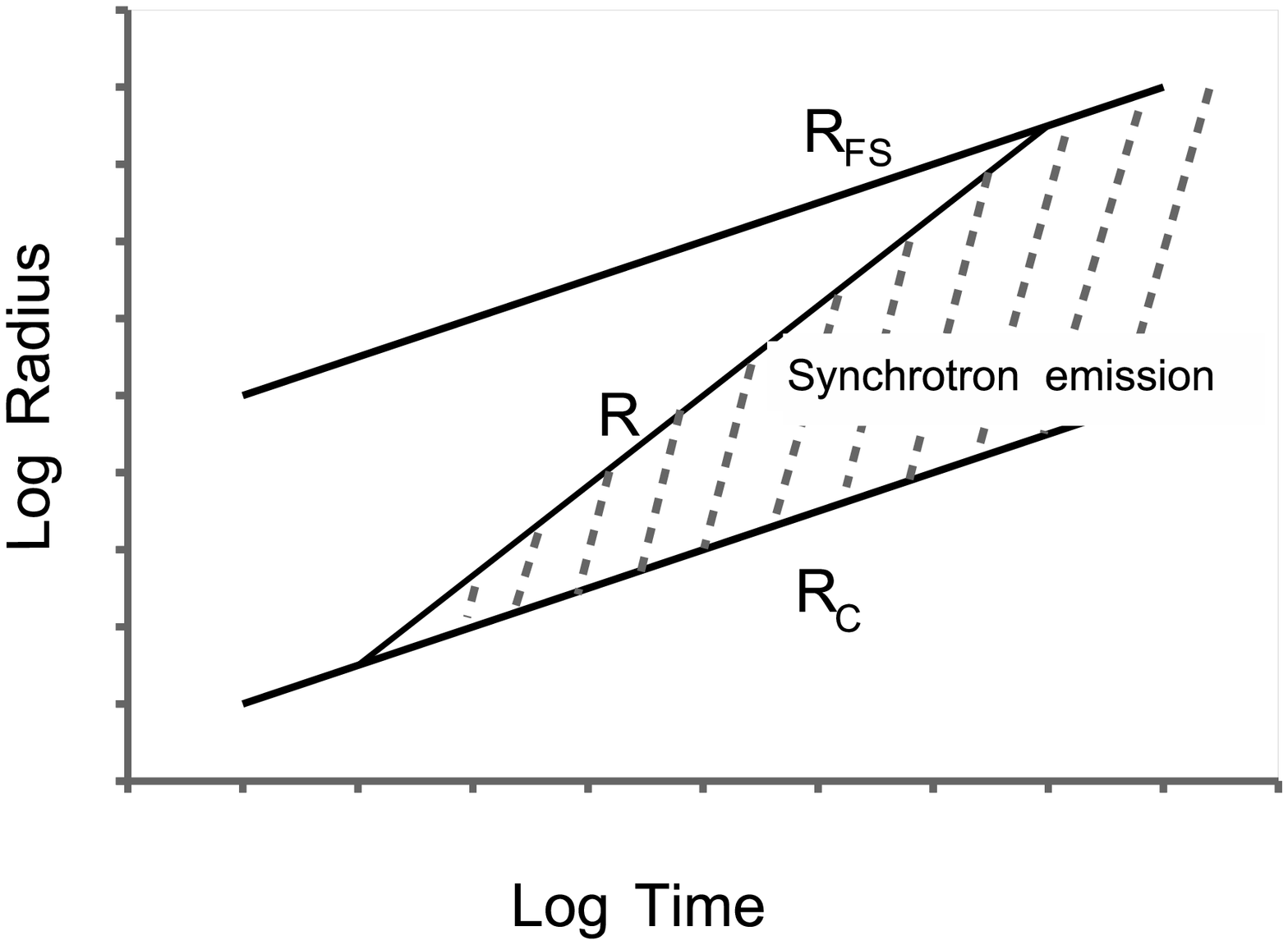}
\caption{A toy model for the evolution of the synchrotron emission region in SN 1993J. $R_{\rm FS}$ and $R_{\rm C}$ are the radii of the forward shock and contact discontinuity, respectively, while $R$ is the outer radius of the expanding emission region. Synchrotron radiation is assumed to come only from the dashed area between $R$ and $R_{\rm C}$. In the figure, it is assumed that the emission region during the later phase extends all the way to the forward shock. As discussed in the text, this is not necessary, only that the outer radius $R$ becomes a constant fraction of $R_{\rm FS}$.\label{fig1}} 
\end{figure}

In previous modeling of the radio and X-ray emission from SN 1993J, it has tacitly been assumed that the velocity deduced from the expanding VLBI-source corresponds to that of the forward shock. In this section, a revised model is discussed in which this is not the case. It takes its starting point in the VLBI-observations, which show the outer radius of the radio source to increase, initially, almost linearly with time with a later transition to a slower expansion \citep{bar02, bie11}. Extrapolating the later expansion back to the earliest VLBI-observations, one finds that the difference between the extrapolated  and measured radii is approximately equal to that expected between the forward shock and contact discontinuity/reverse shock in the self-similar models of \cite{che82a}. A simple toy model to depict such a situation is shown in Figure\,\ref{fig1}. Since the distance between the contact discontinuity and the reverse shock is much smaller than the distance between the forward shock and the contact discontinuity, no distinction is made between contact discontinuity and reverse shock in the discussion below. The radii of the contact discontinuity ($R_{\rm C}$) and forward shock ($R_{\rm FS}$), respectively, are assumed to vary with time according to the self-similar model. The radius $R$ is the outer radius of the synchrotron source; i.e., the synchrotron emission is produced between $R$ and $R_{\rm C}$. The observed transition of the variation of $R$ at around $300$\,days is assumed to correspond to $R \approx R_{\rm FS}$ (or some constant fraction thereof). At later times, the expansion of the VLBI-source is determined by the forward shock. 

Such a model can provide alternative explanations to both the radio and X-ray observations. As discussed in Section\,\ref{sect2}, the initial decrease of the self-absorption frequency ($\nu_{\rm abs}$) with time was slower than expected for an adiabatic source. With synchrotron cooling, the thickness of the synchrotron emitting shell increases faster than for the adiabatic case and observations could be accounted for by standard assumptions regarding the variation of the magnetic field strength and density of relativistic electrons. Hence, with an appropriate choice of $R(t)$, the same result can be obtained for an adiabatic source model.

The roughly constant velocity of the VLBI-source during the first few hundred days implies a large value for $n$ in the standard, self-similar model. As discussed in \cite{fra96}, the reverse shock is then cooling and the observed X-ray emission comes from behind the forward shock. In the proposed revised model, the value of $n$ is not constrained during the initial phase and, for example, it may be the same as that appropriate for the later phase. Hence, the initial VLBI-measurements cannot be used as arguments against adiabatic models. Rather, arguments in favor of adiabatic models can found from two independent aspects of the observations, namely, X-ray emission \citep{suz93} and free-free absorption of the synchrotron emission \citep{b/l14}, both of which give similar values of $\dot M_{\rm -5}$ and $n$. It may be noticed that in order for adiabatic X-ray models to be applicable, it is necessary for the synchrotron emitting region to grow outwards starting from the contact discontinuity rather than inwards starting from the forward shock. 

The rate of increase of the outer VLBI-radius changes at a few hundred days. \cite{bar02} show that a good fit can be obtained with two power-laws, which cross at approximately $300$\,days. If this were due to increased deceleration of the forward shock, one would expect the transition to take place over at least a dynamical timescale. In such a situation, the measured radii during the transition phase should lie below the extrapolations of both power-laws. The observed error bars are small enough to argue that this may not be the case. If so, this would indicate a rather abrupt transition more akin to a discontinuity in the velocity. Such a discontinuity is straightforward to account for in the revised model but harder to incorporate into a model where the increasing deceleration of the forward shock is due to external changes in either the circumstellar matter or the supernova ejecta. 

In principle, the VLBI-observations could be used to deduce $R(t)$. Hence, for a given value of $n$, and assuming a homogeneous distribution of relativistic electrons and magnetic fields in the region $R - R_{\rm C}$, the synchrotron self-absorption could be calculated and compared to the observed $\nu_{\rm abs}(t)$. However, there are several effects that limit the usefulness of such a comparison. As will be discussed in the next section, at least the magnetic field is likely to be highly inhomogeneous in the synchrotron emitting region. Furthermore, during the initial phase, the outer and inner radii of the VLBI-source could not be determined independently. Instead a constant ratio was assumed, which was taken to be the same as that observed during the later phase of expansion. Such an assumption is inconsistent with either synchrotron cooling models or the proposed revised model. Although this is not expected to change the deduced values of the outer radius substantially, it can have a non-negligible effect on $\nu_{\rm abs}(t)$.

In order to illustrate a few of the implications of the revised model, let $n=7$. For a given observed radius, the instantaneous velocity depends on $n$ as $v \propto (n-3)/(n-2)$. Hence, at an assumed transition on day $300$, the velocity of the forward shock is smaller than in the nearly constant velocity scenario by a factor $0.8$, i.e., $v_{\rm sh,4} = 2.2 \times 0.8$. From \cite{che82a}, one also finds that $v_{\rm ej} = 0.93\, v_{\rm sh}$. The variation of the ejecta velocity at the reverse shock is then given by $v_{\rm ej} = 1.6 \times 10^4 (t/300\,{\rm days})^{-0.2}$\,km\,s$^{-1}$. The first radio and X-ray observations took place at $t \approx 10$\,days, which corresponds to an ejecta velocity $v_{\rm ej} \approx 3.2 \times 10^4$\,km\,s$^{-1}$. This is close to the maximum velocity of model $13$B in \cite{woo94} and shows that this model is consistent with a self-similar solution also during the early observations. 

\cite{wei07} have emphasized that an achromatic break occurs in all the radio as well as X-ray light curves at $t \approx 3100$\,days, which corresponds to $v_{\rm ej} \approx 1.0 \times 10^4$\,km\,s$^{-1}$ This value matches closely the ejecta velocity in model $13$B, where the density distribution makes a sharp break and enters a region where it stays almost constant with decreasing radius/velocity. This region represents the inner part of the hydrogen rich envelope. A possible cause for the achromatic breaks in the light curves is, therefore, that the energy/momentum from the ejecta is no longer large enough to maintain a self-similar shock structure; i.e., the reverse shock would weaken or even disappear. Not only would this reduce the X-ray emission but also, as is discussed in the next section, cause a decline in the radio emission. Support for such a scenario comes from the VLBI-observations of \cite{bie11}. After a few years, the radio source had expanded enough for the observations to allow individual measurements of the outer and inner radii of the synchrotron emitting shell. The outer as well as the inner radius follow a $n \approx 7$ evolution until $\approx 3000$\,days. At this time, the expansion of the inner radius slows down considerably, while the outer radius is much less affected. In the revised model, the base of the radio emitting region is the contact discontinuity. A weakening of the reverse shock is expected to affect the velocity of the contact discontinuity and, hence, the velocity of the inner radius of the radio source.

\section{Discussion}\label{sect4}

The toy model shown in Figure\,\ref{fig1} envisions a synchrotron emission region that starts to grow outwards from the contact discontinuity to fill all, or a constant part, of the region behind the forward shock. Such a scenario accords well with the physical situation expected to emerge from the Rayleigh-Taylor instability at the contact discontinuity. Density contrasts by more than a factor $10$ result as shocked ejecta and circumstellar matter mix \citep{che92}. Likewise, the strength of the magnetic field amplified by the turbulence varies considerably, with the highest values in regions bounding the high density fingers of ejecta gas protruding from the contact discontinuity \citep{j/n96a}. Roughly half of the volume between the forward and reverse shocks is estimated to be affected by the instability in its saturated phase. This fraction is likely to increase with the compression ratio across the forward shock; for example, due to a significant amount of relativistic electrons/ions \citep{blo01, duf14}. The physical underpinning of the toy model in Section\,\ref{sect3} suggested here is then similar to the one discussed by \cite{che92} and \cite{j/n96a} for the radio emission in SN 1572 (Tycho) not directly associated with the forward shock. The association of the synchrotron source with the turbulent region implies two things. Firstly, the amplified magnetic field should dominate the magnetic field behind the forward shock \citep[e.g.,][]{j/n96b}. Secondly, in order for the synchrotron emission region to be defined by the magnetic field distribution, the relativistic electrons should fill this region more or less uniformly; for example, through first order Fermi-acceleration at the forward shock. 

The effects of the amplification of the magnetic field by the Rayleigh-Taylor instability on the spatially resolved VLBI-source have also been considered in \cite{bie03}. The focus was on the evolution of SN 1993J after a few years, when the outer and inner radii could be resolved individually and the source was in the strongly decelerating phase. They concluded that the outer radius is rather close to the forward shock but find no indications for the shock-front itself to be affected by turbulence. In contrast to the revised model discussed in this paper, they suggested that the two different velocity regimes may indicate that the shock structure was not self-similar. 

The radio observations of SN 1993J clearly showed the brightness temperature to be below that expected for a homogeneous synchrotron source in which the magnetic field is in equipartition with the relativistic electrons. The slower than expected decrease of the synchrotron self-absorption frequency increased the brightness temperature with time. However, in \cite{f/b98} the brightness temperature is also lowered by the dominance of the magnetic energy density over that in relativistic electrons, which provides a time independent factor. In the revised model, the initial increase of the brightness temperature and the associated slow decrease of the synchrotron self-absorption frequency are due to the relative increase of the thickness of the synchrotron emitting shell, while the constant factor is attributed to the inhomogeneous structure of the magnetic field, which results in an effective covering factor smaller than unity \citep{bjo13}. A more quantitative description of the variations of brightness temperature and synchrotron self-absorption frequency is harder to do, since this is sensitive to the details of the inhomogeneous source structure.

As discussed in Section\,\ref{sect3}, the break in the X-ray light curves at $t \approx 3100$\,days may be due to the reverse shock entering a roughly flat portion of the density distribution of the ejecta, which makes the self-similar solutions inapplicable. With the revised model discussed above, the simultaneous break in the radio light curves can be ascribed to the same physical effect. As discussed in \cite{j/n96a}, a weakening of the reverse shock causes the density contrast at the contact discontinuity to decrease. This, in turn, weakens the driving of the Rayleigh-Taylor instability and, hence, leads to less amplification of the magnetic field. The accompanying increase of the region behind the reverse shock is then the likely reason for the slowing down of the expansion of the inner boundary observed by \cite{bie11} at the same time. 

Some of the aspects of the revised model are also conducive to limiting the possible interpretations of the X-ray observations. It was argued in \cite{fra96} that the low temperature deduced for the X-ray emitting gas around $200$\,days \citep{zim94} is hard to accommodate within an adiabatic model. However, later observations have indicated \citep{z/a03, swa03} that there is a range in temperatures. Although a consistent description has not yet been achieved, the implied range is at least a factor of $10$. \cite{z/a03} fitted the X-ray emission up to $ \approx 3000$\,days with a simple two temperature model, which indicates that the average temperature decreases with time but that the range is more or less constant. In light of the revised model, it is interesting to note that the evolution of both temperatures is consistent with equipartition between ions and electrons and $n\approx 7$. A similar conclusion has been reached by \cite{uno02} when fitting the {\it ASCA} observations during the first $\approx 600$\,days.

Such a range in temperatures may be caused by the turbulence driven by the Rayleigh-Taylor instability at the contact discontinuity. When the instability has saturated, it is likely to cover a substantial part of the shocked circumstellar medium. In the homogeneous, adiabatic model, the X-ray emission should come mainly from the shocked ejecta. However, the large density variations in the turbulent region could result in a non-negligible contribution of the shocked circumstellar medium to the X-ray emission even in the case of an adiabatic reverse shock. Although the relative contribution to the total X-ray emission from the two shocked regions is rather sensitive to the details of the revised model, there are a few aspects that distinguishes it from homogeneous models with an initially nearly constant shock velocity (i.e., high value of $n$).

The higher shock velocities early on in the revised model as compared to constant velocity models increase the equipartition temperatures behind the shocks; for example, assuming equal numbers of hydrogen and helium atoms \citep{bar94}, the equipartition temperature behind the reverse shock is $\approx 10^9$\,K at $10$\,days. Although, in order for the electrons to reach this temperature, mechanisms in addition to Coulomb collisions are needed to transfer energy from the ions to the electrons, it shows that the reverse shock may have contributed to the high energy emission detected by OSSE \citep{lei94}. The difference between the equipartition temperature and the temperature reached by the electrons behind the reverse shock, assuming Coulomb collisions only, is not that large. The opposite applies for the electrons behind the forward shock, where this difference is much larger. The turbulence driven by the density contrast at the contact discontinuity amplifies the magnetic field. This process may also excite plasma waves, which could contribute to the energy transfer between ions and electrons. If so, this would further enhance the relative importance of the forward shock to the X-ray emission.

When the electron temperature ($T_{\rm e}$) is below its equipartition value and determined by Coulomb collisions, it varies as $T_{\rm e} \propto (n_{\rm e} T_{\rm ion} t)^{2/5}$ \citep{spi68}, where $T_{\rm ion}$ is the ion temperature. Since $T_{\rm ion} \propto v_{\rm sh}^2$, in a steady wind environment $T_{\rm e} \propto t^{-2/5}$ is expected. This scaling relation coincides with that for the equipartition temperature when $n=7$. The deduced temperature variations discussed above may then not necessarily indicate that equipartition conditions apply.

The observed X-ray emission limits the contribution from Comptonized flux. This leads to a tension between the high temperature indicated by the OSSE-observations and the high mass-loss rates needed to account for the softer X-ray emission. As is seen in \cite{fra96}, this causes considerable strain when trying to find acceptable parameters in radiative models. This constraint is eased in adiabatic models; for example, $\dot M_{\rm -5} = 1$ and $n = 7$ give an optical depth to electron scattering around $10$\,days that is more than a factor $10$ smaller than for the best fit model in \cite{fra96}. This then allows for correspondingly higher temperatures.

\section{Conclusions}\label{sect5}

The main conclusion of the present paper is that a revision is needed of the model(s) used to describe the observed behavior of SN 1993J. This is based on the following two main results from the analysis in Section\,\ref{sect2}.

1) Together, radio and X-ray observations show that there are a number of arguments against models invoking cooling in either spectral regime; for example, the implied mass and energy at high eject velocities are at least an order of magnitude larger than expected in standard core collapse scenarios.

2) Adiabatic models can account for the radio as well as X-ray observations; for example, a mass-loss rate of the progenitor star around $10^{-5} M_{\rm \odot}$\,yr$^{-1}$ (for a wind velocity $v_{\rm w} = 10$\,km\,s$^{-1}$) and an ejecta density distribution corresponding to $n\approx 7$ are consistent with both radio and X-ray observations.

The crucial new component in the revised model is the interpretation of the radius of the spatially resolved VLBI-source.

3) During the first few hundred days, when the velocity of the radio supernova is roughly constant, the observed radius does not correspond to that of the forward shock. Instead, it is associated with the expansion of the Rayleigh-Taylor unstable region starting from the contact discontinuity. After saturation of the instability, the outer radius of the VLBI-source is determined by the forward shock. Hence, the velocity of the forward shock is determined by $n \approx 7$ also during the initial phase of evolution of the supernova.

4) The radio emission results from the amplification of the magnetic field by the turbulence in the Rayleigh-Taylor unstable region. Such an explanation has previously been put forth for the radio emission sometimes observed to peak inside the forward shock in supernova remnants, for example, SN 1572 (Tycho). Hence, SN 1993J would offer the possibility to study also the time evolution of this instability.

5) A low value of $n$ makes it easier to determine whether the value of a given quantity scales with radius ($R$) or time ($t$). In an adiabatic model for SN 1993J with $n \approx 7$, observations suggest that the magnetic field ($B$) varies as $B \propto 1/R$ rather than $B \propto 1/t$; i.e., the energy density of the magnetic field does not scale with the thermal energy density behind the forward shock. Furthermore, with $B \propto 1/R$, observations indicate that also the energy density of relativistic electrons does not scale with the thermal energy density.

6) Both the ejecta mass and velocity structure deduced for SN 1993J accord well with model $13B$ favored by \cite{woo94} from an analysis of the optical observations.

\clearpage

\end{document}